# Resource and Competence (Internal) View vs. Environment and Market (External) View when defining a Business

Yngve Dahle
Norwegian University of Science and Technology (NTNU)
Faculty of Engineering Science and Technology (IVT)
Norway
yngve.dahle@ntnu.no

Martin Steinert
Norwegian University of Science and Technology (NTNU)
Faculty of Engineering Science and Technology (IVT)
Norway
martin.steinert@ntnu.no

Anh Nguyen Duc
University College of Southeastern Norway
Norway
anh.nguyen.duc@usn.no

Roman Chizhevskiy
Lean Business AS
Norway
roman@leanbusiness.no

*Abstract* — **Startups is a popular phenomenon that has a significant impact on global economy growth, innovation and society development. However, there is still insufficient understanding about startups, particularly, how to start a new business in the relation to consequent performance. Toward this knowledge, we have performed an empirical study regarding the differences between a Resource and Competence View (Internal) vs Environment and Market View (External) when defining a Business. 701 entrepreneurs have reflected on their startups on nine classes of Resources (values, vision, personal objectives, employees and partners, buildings and rental contracts, cash and credit, patents, IPR's and brands, products and services and finally revenues and grants) and three elements of the Business Mission ("KeyContribution", "KeyMarket" and "Distinction"). It seems to be a tendency to favour the Internal View over the External View. This tendency is clearer in Stable Economies (Europe) than in Emerging Economies (South Africa). There seems to be a co-variation between the tendency to favour the Internal View and the tendency to focus on adding Resources. Finally, we found that an order-based analysis seems to explain the differences between the two views better than a number-based method.**

*Keywords*: *Resource Based Model, Market Based Model, Knowledge Based Model, Environmental Models of Competitive Advantage, Internal View, External View, Distinction, Key Market, Key Contribution, Big Data, Business Mission, Business Idea, Data Warehouse, Lean Startup,*

I. INTRODUCTION AND THEORY

Starting a new venture is a common trend, especially at the modern world with advanced development and adoption of technology. There is a rapid growth of products and services created by newly created, little to none historical operation companies that attempts to build and scale their businesses. But first and foremost, one would be interested in the question -what is the best way of starting and developing a Startup? [1] Entrepreneurship literature offers two alternative school of thoughts on this question. While the first focuses on leveraging existing resources and competences, the second focuses on adapting as good as one can to market conditions regarding customers' identity and needs, competitors, substitutes and suppliers.

J. Barney has divided the approach of understanding the sustained competitive advantage of a firm into two different viewpoints, namely the *Resource Based Model* and the *Environmental Models of Competitive Advantage* [2]. The *Resource Based Model* examines the relationship between a firms' internal weaknesses and strengths (Internal Analysis) and its performance. These resources, according to Barney; consists among other things of the firms "assets, capabilities, organizational processes, firm attributes, information and knowledge". Wernerfelt has defined resources as "anything that could be thought of as a strength and a weakness of a given firm. More formally, a firm's resources at a given time could be defined as those (tangible and intangible) assets that are tied semi-permanently to the firm." [3]. Prahalad and Hamel has extended this in their discussion of the "Competence Based View" [4]. This theory enhances the Resource Based Model by focusing further on the Resources that constitutes "Unique Knowledge" in the organization. They claim that the competitive advantage of a firm is better understood by researching the core competences behind products than researching the products itself.

For the purpose of this article, we are not going to separate between the Resource Based Model and the Competence Based View. For the remainder of this article we will simply call this viewpoint the *Internal Viewpoint*.

The *Environmental Models of Competitive Advantage,* on the other hand, tries to understand competitive advantages by primarily analyzing the organization's external opportunities and threats. This latter model draws heavily on Porter's Five Force Model [5], where the company is understood in accordance to a competitive ecosystem containing threats from new entrants and substitutes, the bargaining power of suppliers and customers and finally, the existing competitors within the industry. For the remainder of this article we will call this viewpoint the *External Viewpoint*.

Our research objective is to investigate the adoption of internal and external viewpoints among the Startup population. Derived from this objective, we present the Research Question (RQ):

*RQ: How do Startups perform their business development regarding to the Internal and External viewpoint?*

We argue that there may be an interesting contribution to this debate by adding Christopher K. Bart's [6] definition of the *Business Mission* to the analysis. According to Bart, any firms core identity can best be understood by looking at the interaction of three core phenomena in parallel: "KeyContribution" (what problem do you solve, or what need do you satisfy), "KeyMarket" - what variables does best describe the firms typical customers and "Distinction" - which describes the unique competence that enables the firm to be competitive in solving this problem.

If we combine the separation into the Internal and External Viewpoint with the Business Mission thinking, one could imagine that a firm with a primary tendency to view themselves according to the Internal Viewpoint, would have a higher propensity to focus on the Internal (or Competence) Element in the Business Mission, namely "Distinction". Similarly, one could think that a firm with a primary tendency to view themselves according to the External Viewpoint, would have a higher propensity to focus on the two External Elements in the Business Mission, namely "KeyContribution" and "KeyMarket". Furthermore, one could imagine that a firm with a strong emphasis on defining and developing their Resources would tend to have a higher tendency to have an Internal Viewpoint.

Now we need to find a way to measure these parameters. We can use our Entrepreneurial Data Warehouse (EDW) to take a look into this. The construction of the EDW draws heavily on Blank, Ries and Maurya [7,8,9].

This paper is organized as follows, Section II describes the Entrepreneurial Data Warehouse. Section III introduces three new theoretical constructs that we have made to simplify the understanding of our data. Section IV describes the way the empirical test is set up. Section V presents the empirical findings. Section VI discusses weaknesses and problems with the design. Section VII contains the conclusions.

## II. THE PLATFORM AND THE ENTREPRENEURIAL DATA WAREHOUSE

Since the 15th of November 2017, we have had our Entrepreneurial Data Warehouse fully operative [10]. So – this paper is the first attempt to utilize the data in the EDW. The fact that the EDW has been operational for such a short time suggest that the data will be more powerful at a later stage. Between November 15th, 2017 and March 1st, 2018 – there have been added a number of 701 new cases relevant to this study to the platform, where each entrepreneur has been adding a plethora of information about their businesses over the time span. Among other things, the entrepreneurs have added their *Resources* and their *Business Mission*. This Section of the paper will describe how the Entrepreneurship Data Warehouse is set up, and how it can help us shed light on the Internal versus the External Viewpoint:

What set our project apart, is that we are creating a uniform categorization of the different theoretical events that constitutes a business development process. We call these *Event Categories*. The Entrepreneurship Platform (EP), as shown in Figure 2, is the tool that enables us to monitor the Events created, updated and deleted within each Event Category by a large number of entrepreneurs [11].

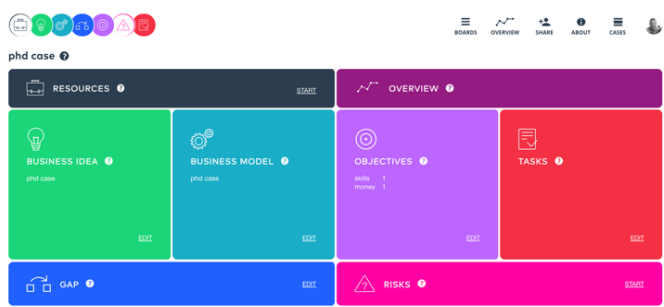

*Figure 1: The Entrepreneurial Platform*

The EP consists of 7 Boards, 3 different Customer Test Features and an Admin Console. Each of these consists of a number of *Boxes*, and the User Interface is based on filling these boxes with *Cards* containing the relevant Content. Everything is linked together, resulting in an *Overview*, giving the Case Company control over their current strategy, project and forecast. The thinking behind this model is clarified in the book Lean Business Planning (Dahle, Dagestad, Alskog, & Bang Abelsen, 2014)

The EP is marketed directly as a free-of-cost SaaS tool toward startups. In addition, it is marketed in white-labelled versions from a number of Banks, Innovation Centres, Venture Capitalists and Incubators.

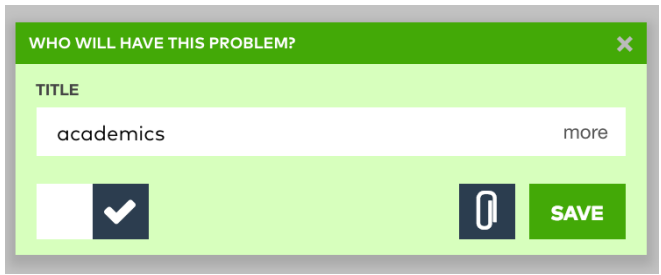

*Figure 2: Example of Card. One of the Business Idea Cards*

In this article, we will focus on only two of the seven boards in the system, the *Resource Board* and the *Business Idea Board*. The content from all other boards, will be ignored.

The *Resource Board* have nine boxes (as shown in Figure). The first box gives the different stakeholders a chance to define the *Values* they want to base the business on. The second box fits Cards that suggest the *Vision* of the business. A well-conceived vision consists of two major components: core ideology and envisioned future. The core ideology is unchanging while the envisioned future is what we aspire to become, to achieve, and to create [12]. In the third box, the different owners of the Case Company can enter Cards with their personal *Objectives* for the Case Company. The fourth box contains cards for *Employees and Partners of the firm*. The fifth box is *Buildings and Rental Contracts*, in case the firm have one or several building or rental offices. The sixth box is about *Cash and Credit*, giving the detail of financial situation of the firm. The seventh box registers *Patents, Intellectual Properties and Brands* that the firm assesses. The eighth box presents cards for each *Products and Services* of the firm. And finally, the ninth box describes *Revenues and Grants*.

*Figure 3: The Resource Board*

The *Business Idea Board* (Figure 4) allows the stakeholders to define their *"Business Mission"*. As previously mentioned (Bart, 1997), the Business *Mission* consist of three components:
1. "KeyContribution": What problem will you solve?
2. "KeyMarket": Who will have this problem?
3. "Distinction": What makes you unique?

Each Entrepreneur will first populate the different Boxes with Cards in a brainstorming type process. The Companies are then advised to link the different Cards to as many different Business Ideas as they please. We will ignore this latter process in this article.

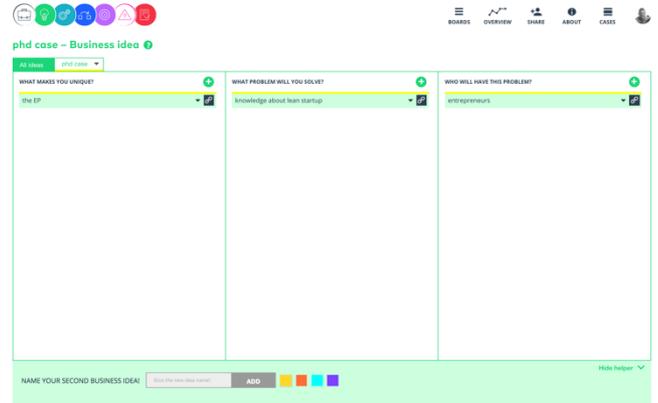

*Figure 4: The Business Idea Board*

Thus, for this article, we will focus on 12 Event Categories, as shown in Table 1. The source of Event can be the Resource board (RES) or the Business Idea board (BI).

*Table 1: Event Categories*

| # | Source | Event Category |
|---|--------|----------------|
| 1 | RES | Values |
| 2 | RES | Vision |
| 3 | RES | Owner's Objectives |
| 4 | RES | Employees and Partners |
| 5 | RES | Buildings and Rental Contracts |
| 6 | RES | Cash and Credit |
| 7 | RES | Patents. IPR's and Brands |
| 8 | RES | Products and Services |
| 9 | RES | Revenues and Grants |
| 10 | BI | Key Contribution |
| 11 | BI | Key Market |
| 12 | BI | Distinction |

### III. INTRODUCING THREE NEW CONSTRUCTS

The amount of data allows us to perform various types of quantitative analysis. The first step is to define metrics that are relevant and meaningful for our given theories. To be able to make a meaningful analysis, we will simplify the vast amount of information in the EDW into three simple quantitative constructs, namely The Resource Count, The Number Coefficient and The Order Coefficient

**A. The Resource Count**. First, we want to simply *count the number of Resource Card*s defined for each Case. We call this the *Resource Count*, as described in the formula below:

$TheResourceCount = $ #Values + #Vision + #Owner's Objectives + #Employees and Partners + #Buildings and RentalContracts + #Cash and Credit + #Patents.IPR's and Brands + #Products and Services + #Revenues and Grants.

For the Resource Count, we have excluded all South African Cases, as the system in South Africa has included only three of the nine Resource Boxes. This has left us with 281 Non-South-African cases. In these cases, a total of 3317 Resources have been entered, averaging 11,8 Resources per Case. The averages have been distributed as follows over the nine Resource Classes, as shown in Table 2

*Table 2: Resource Count Averages*

| Resource element | All (excluding Afrian cases) |
|---|---|
| Core Values | 2,44 |
| Visions | 1,44 |
| Owners Objectives | 1,84 |
| Employees & Partners | 1,37 |
| Buildings & Machines | 0,73 |
| Cash & Credit | 0,52 |
| Patents, IPR and Brands | 0,51 |
| Products | 2,25 |
| Revenues and Grants | 0,70 |

Next, we want to create two different constructs for the propensity for an Internal Viewpoint versus an External Viewpoint in a case, based on the relationship between the "Distinction" cards, the "KeyContribution" cards and the "KeyMarket" cards. These coefficients are calculated for all the 701 cases in the sample.

### B. The Internal Viewpoint/External Viewpoint *Number Coefficient*

The Number Coefficient (NC) as we will call it here, is calculated as follows:

$$NC = \frac{2 \,(\text{\# of "Distinction" cards})}{(\text{\# of "KeyContribution" cards} + \text{\# of "KeyMarket" cards})}$$

The Number Coefficient has a theoretical range from 0 to +∞ (in the very theoretical case that the entrepreneur enters an unlimited number of distinction cards). The numerator represents the Internal View and the denominator represents the External View. Any Coefficient higher than +1 suggests a propensity for the Internal View. Any Coefficient between 0 and +1 suggests a propensity for the External View.

### C. The Internal Viewpoint/External Viewpoint *Order Coefficient*

Another way of looking at the Internal Viewpoint versus External Viewpoint propensity, is to analyze the *order* in which the cards are entered. Since the entrepreneurs are allowed to enter the cards in any order they like, it could be possible to assume that the elements most important for the entrepreneurs would be entered first. We base this in an exponential simulation, where we have assigned the value 200 to the first card entered, and then 150, 125, 100, 80, 60, 40, 20, 10 and 5 for the next 9. Cards after the 10$^{th}$ have been given no value. The Formula for the Order Coefficient (OC) is

$$OC = \frac{2\Sigma(D^1, D^n)}{\Sigma(K^1, K^n)}$$

If "Distinction" is represented by D, and both "KeyContribution" and "KeyMarket" is abstracted by K, one example of the order of entry is:

D → D → D → D → K → D → K → K → K → K
200 → 150 → 125 → 100 → 80 → 60 → 40 → 20 → 10 → 5

The calculation of values of D and K in this example, will be as follows:

D=200+150+125+100+60= 635
K=80+40+20+10+5=155
D/K= 635/155
D/K= 4,10

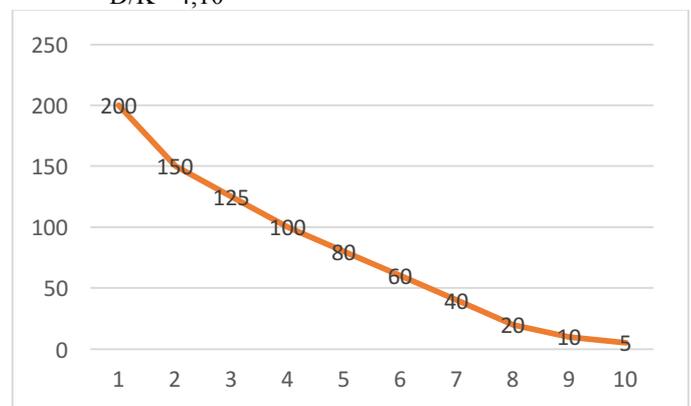

*Figure 5: The Order Coefficient Exponential Curve*

The Order Coefficient has a range from 0 to +314 (2*(200+150+125+100+80+60+40+20+10)/5). This represents the case that an entrepreneur enters only one K-card – and that is the 10$^{th}$ and last card entered. The numerator represents the Internal View and the denominator represents the External View. Any Coefficient between +1 and +314 suggests a propensity for the Internal View. Any Coefficient between 0 and +1 suggests a propensity for the External View.

## IV. THE RESEARCH DESIGN

First of all, we have to remember that this is an early attempt to utilize the EDW. So – we will not attempt to make any novel conclusions based on the material, but rather try to shed some light on the Internal View versus External View Axis. We still have a very limited number of observations for a quantitative study. This is rapidly increasing for every day that passes. Thus – we can at a later point extend this study both with regard to the number of observations and by looking more closely at the qualitative content of each case. The estimates are that the EDW will surpass 10.000 cases by the end of 2018.

The starting point for this research, is the 701 Cases in the EDW registered between November 15th, 2017 and March 1st, 2018, that are defined as real live cases. All test, demo cases and incomplete cases are removed in a manual data pre-processing step. Altogether the accepted cases had 6825 Resource Cards and 8754 Business Mission Cards registered. For the analysis of the co-variation between the Resource Count and the NC and OC, we disqualify all the South-African Cases – as these have been presented with a simplified Resource Board. This limited sample include 281 cases, 3317 Resource Cards and 4036 Business Mission Cards. In this analysis, we have only included Create and Update Events, as Delete Events are deemed irrelevant. Finally, 678 of the 701 Cases come from five countries, as shown in Figure 6, which are South Africa (420 cases), Norway (140 cases), Denmark (56 cases), Sweden (34 cases), and United Kingdom (28 cases).

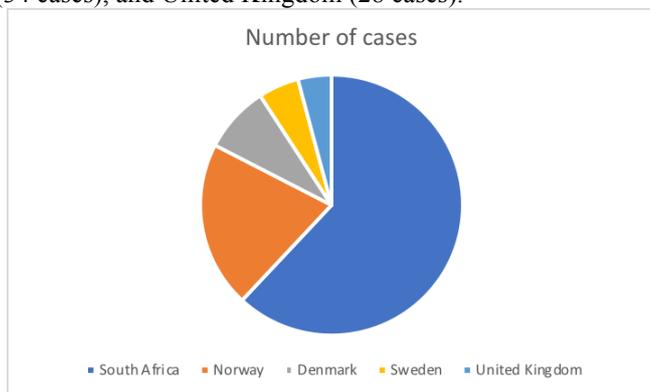

*Figure 6: Country Distribution*

To shed light on the concept of the Internal View versus the External View, we break down our main RQ into three sub questions (SQs)

***SQ1: Is there generally, for startups, a higher tendency toward the Internal View in comparison to the External View?***
One could imagine that firms with a primary tendency to view themselves according to the Internal Viewpoint, would have a higher propensity to have a high focus on their Resources. Therefore, one would think that any case with a high NC or OC would also tend to have a high Resource Count. Here we have separated the Sample into three parts, separately for the Number Coefficient and the Order Coefficient. The Higher Quartile represent the 25% of the cases with the highest NC or OC. The Lower Quartile represent the 25% of the cases with the lowest NC or OC. The remaining 50% in the middle is not taken into consideration. Thereafter we have analysed the Co-variation between the Resource Count and the NC and OC in the Higher Quartile and the Lower Quartile respectively. We come up with a list of hypotheses to explore We come up with a list of hypotheses to explore this SQ1: SQ1:

- H1a: Null hypothesis: There is total neutrality between the Internal and the External View, thus the NC would be 1.0
- H1b: Alternative hypothesis: There is an Internal View leaning tendency in startups (NC>1)
- H2a: Null hypothesis: There is total neutrality between the Internal and the External View, thus the OC would be 1.0
- H2b: Alternative hypothesis: There is an Internal View leaning tendency in startups (OC>1)
- H3a: Null hypothesis: There is no difference in the Resource Count between lower quartile NC and higher quartile NC.
- H3b: Alternative hypothesis: There is a statistical difference in the Resource Count between lower quartile NC and higher quartile NC.
- H4a: Null hypothesis: There is no difference in the Resource Count between lower quartile OC and higher quartile OC.
- H4b: Alternative hypothesis: There is a statistical difference in the Resource Count between lower quartile OC and higher quartile OC.

We would perform a one tail t-test for mean value of NC and OC values to test these hypotheses.

***SQ2: Is there a geographical difference in Startups' tendency toward the Internal View or the External View?***
Entrepreneurship in different geographical areas, tend to be very culturally different [13]. In particular, these differences are clear between emerging economies and more stable entrepreneurial ecosystems. Is this also true for the Internal View versus External View Axis? We will check this by comparing the NC and OC of the African cases versus the European Cases. We come up with this hypothesis to explore SQ2:

- H5a: Null hypothesis: There is total neutrality regarding the difference between the Internal and the External View in the African cases versus the European Cases.
- H5b: Alternative hypothesis: There is higher propensity for the Internal View in the European cases than in the African Cases.

***SQ3: Is there a difference between the Number Coefficient and the Order Coefficient in describing SQ1 and SQ2 ?***
Finally, it will be interesting to see which one of the NC and the OC (if any) that proves to give the clearest results. Whether the sheer *number* of "Distinctions" versus "KeyContributions" and "KeyMarkets" will be the most useful concept, or whether the order in which these elements are added into the case gives more insight. We can loosely discuss this by looking at how the two concepts differ in the first two SQs. We do not expect to come up with hard evidence on any terms usefulness, but hopefully we will be able to derive some interesting learning points from it.

- H6a: Null hypothesis: There is total neutrality between the explanation power of the Number Coefficient and the Order Coefficient.
- H6b: Alternative hypothesis: The Order Coefficient has a higher explanation power than the Number Coefficient.

## V. FINDINGS

### A. General Tendency toward Internal versus External View (answering SQ1)

Both the Number Coefficient and the Order Coefficient is designed in such a way that a complete balance between the Internal View (Focus on the "Distinction") and the External View (Focus on "KeyContributions" and "KeyMarkets") will give a coefficient of 1.0. Any Coefficient higher than 1.0 will suggest a higher focus on the Internal View and any Coefficient below 1 will suggest a higher focus on the External View.

The average Number Coefficient for the 701 cases, is 1.2, with a standard deviation of 0.85 at 99% statistical significance.

```
t = 37.289, df = 700, p-value < 2.2e-16
alternative hypothesis: true mean is
greater than 1
99% confidence interval: (1.126123, Inf)
Mean value: 1.200571
```

We accept the alternative hypothesis H1b here. This shows a slight tendency toward the Internal View.

The average Order Coefficient for the 701 cases, is 4.89, with a standard deviation of 16.7 at 99% statistical significance.

```
t = 7.7439, df = 700, p-value = 1.693e-14
alternative hypothesis: true mean is
greater than 1
99 percent confidence interval:(3.41795,
Inf)
mean of x: 4.886163
```

We accept the alternative hypothesis here. This shows a statistically significant tendency toward the Internal View. This supports Barney [2], who claims that the tendency toward an Resource based (Internal View) is the prevailing tendency.

### B. Covariation between Resource Count and NC/OC (answering SQ1)

To enhance our findings, we will analyze two further things: The definition of Assets, Resources or strengths/weaknesses is rather vague in the Resource Based vs Environmental Based literature [2]. Here we want to study the variations in the Correlation between the Resource Count and the tendency for Internal View versus External View depending on which Resource Event Categories we include. We start with the Number Coefficient, as shown in Table 3

*Table 3: Resource Count per Quartile – Number Coefficient*

| Order Coefficient (OC) | Higher Quartile | Lower Quartile |
|---|---|---|
| Core Values | 2,55 | 2,49 |
| Visions | 1,41 | 1,50 |
| Owners Objectives | 1,71 | 2,11 |
| Employees & Partners | 1,59 | 1,29 |
| Buildings & Machines | 0,73 | 0,58 |
| Cash & Credit | 0,47 | 0,47 |
| Patents, IPR and Brands | 0,45 | 0,40 |
| Products | 2,03 | 2,28 |
| Revenues and Grants | 0,87 | 0,56 |

The distribution of resource count value per quartiles of NC can be seen from Figure 7:

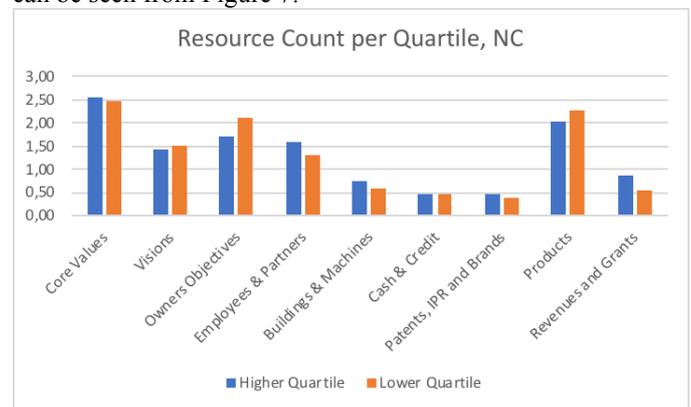

*Figure 7: Resource Count per Quartile, NC*

As in the previous paragraph, the NC do not give very clear answers. Counterintuitively, we see that some of the main Resource Classes have a higher Count for the Lower Quartile than the higher Quartile. The average Resource Count of the Number Coefficient lower quartile cases is 11,13 while the higher quartile cases is 12,54, showing no statistical significance.

```
data:  SUM by NC
t = -1.4976, df = 279, p-value = 0.1354
alternative hypothesis: true difference in
means is not equal to 0
99 percent confidence interval:
 -3.866877  1.035828
sample estimates:
mean in group LQ mean in group UQ
        11.12925         12.54478
```

Here, we accept the Null hypothesis H3a.

Next, we look at the distribution of the Resource Count in the different classes based on the Order Coefficient Quartiles.

*Table 4: Resource Count per Quartile – Order Coefficient*

| Order Coefficient (OC) | Higher Quartile | Lower Quartile |
| --- | --- | --- |
| Core Values | 2,85 | 2,14 |
| Visions | 1,86 | 1,12 |
| Owners Objectives | 2,20 | 1,71 |
| Employees & Partners | 1,49 | 1,45 |
| Buildings & Machines | 0,85 | 0,58 |
| Cash & Credit | 0,58 | 0,54 |
| Patents, IPR and Brands | 0,45 | 0,43 |
| Products | 2,61 | 1,72 |
| Revenues and Grants | 0,58 | 0,69 |

The distribution of resource count value per quartiles of OC can be seen from Figure 8:

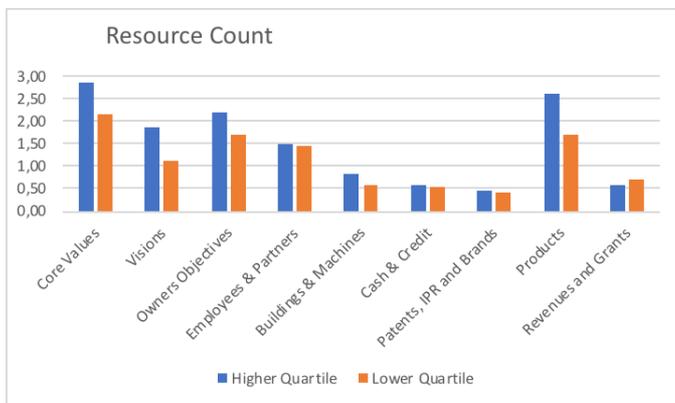

*Figure 8: Resource Count per Quartile, OC*

Here we can see that most of the Resources Categories are quite a lot higher in the Higher Quartile than in the Lower Quartile. The average Resource Count for the Higher Quartile cases are particularly higher for the Core Values, Visions, Owner's Objectives and Products. The average Resource Count of the Order Coefficient lower quartile cases is 10.33 while the average Resource Count for the higher quartile cases is 12.46, at 95% statistical significance.

```
data:  SUM by OC
t = -2.088, df = 279, p-value = 0.03771
alternative hypothesis: true difference in
means is not equal to 0
99 percent confidence interval:
 -4.777635  0.515977
sample estimates:
mean in group LQ mean in group UQ
        10.32558         12.45641
```

Here, we accept the Alternative hypothesis H4b. The defined Resources are very central strategic assets for any entrepreneurial firm, and if a company have a Resource based View, it would make sense that they have a higher focus on defining the most important Resources.

### C. Variations in OC/NC between Emerging and Stable Economies (answering SQ2)

If we split the Number Coefficient up into the five main countries we have in the sample, we get the following results:

*Table 5: Number Coefficient per Country*

| Countries | Number Coefficient |
| --- | --- |
| South Africa | 1,20 |
| Norway | 1,22 |
| Denmark | 1,26 |
| Sweden | 0,99 |
| UK | 1,34 |

Graphically, the distribution of Number Coefficient values across countries are presented in Figure 9.

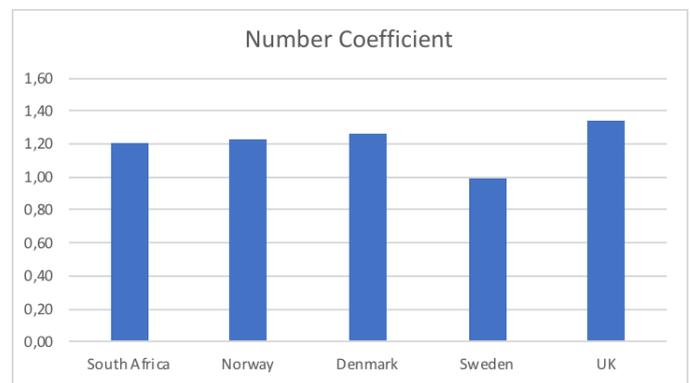

*Figure 9: number Coefficient per Country*

We see that there are no significant findings here. This is in line with other findings with regard to the Number Coefficient. Next, we look at the Order Coefficient:

*Table 6: Order Coefficient per Country*

| Countries | Order Coefficient |
| --- | --- |
| South Africa | 3,62 |
| Norway | 7,56 |
| Denmark | 7,81 |
| Sweden | 2,95 |
| UK | 9,36 |

Graphically, the distribution of Order Coefficient values across countries are presented in Figure 10.

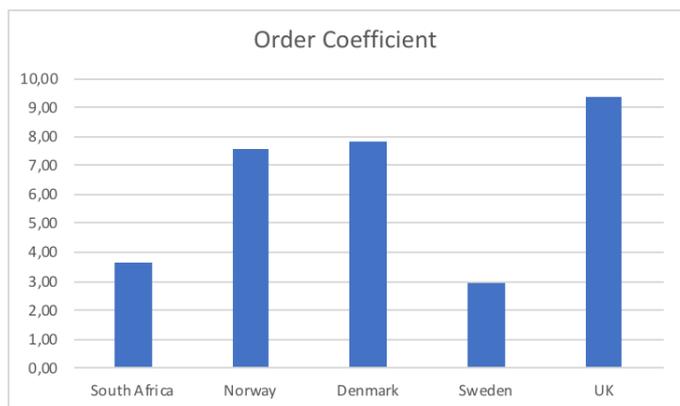

*Figure 10: Order Coefficient per Country*

Here we can see a clear trend. If we break it down into the one African country (South Africa), which we define as an Emerging Economy, and the four European countries, defined as Stable Economies, the difference in the Order Coefficient will be:

*Table 7: Order Coefficient – Africa vs Europe*

| Continents | Order Coefficient |
|---|---|
| Africa | 3,62 |
| Europe | 7,02 |

```
data:  OC by Continent
t = -2.6093, df = 688, p-value = 0.00927
alternative hypothesis: true difference in
means is not equal to 0
99 percent confidence interval:
 -677.952331   -3.436449
sample estimates:
mean in group Africa mean in group Europe
              3.62    7.02
```

Here, we accept the Alternative hypothesis H5b. We see a clearly higher tendency toward the Internal View in Europe than in Africa. The difference is statistical significant at p 0.01. If we define Africa as an emerging market and Europe as a stable market, this has some logical merit. In an unsettled market where there are higher demand and fewer qualified competitors (like Africa), the tendency to focus more on the External market elements would be logical. Similarly, the tendency to focus on one's own "Distinctions" would be logical in a more mature market with more competitors fighting for the customers.

### D. *The differences between the explanation power of the NC and the OC (answering SQ3)*

We see through the analysis that the Order Coefficient tends to give results more in line with what we would logically expect. It gives significant results on all three analyses.

The Number Coefficient does not give a significant result on two of the three questions. This suggest that we should accept the Alternative hypothesis H6b.

This Coincides with a way of thinking that would suggest that the Order in which you add the Business Mission Elements will be a better indicator on the value you put in them than the sheer number of Elements entered.

## VI. THREATS TO VALIDITY

We are building an information gathering setup that is designed to gather a large amount of time series information from tens of thousands of Companies all over the world. There will be some methodological challenges to this operation. Our previous paper [10] discusses many challenges regarding to the study design, data collection and analysis. We will discuss two issues that are specifically linked to this work:

### A. *The "left to right" dilemma*

Due to the practical design of a computer screen, the 9 boxes in the Resource Board and the three Boxes in the Business Mission Board of the EDW will have to be placed according to each other in one or another order. In the Business Mission Board, the "Distinction" is placed to the left, the "KeyContribution" is placed in the middle and the "KeyMarket" is placed to the left. If you are used to write from left to right (as we are in the western world), this *could* cause bias when considering the Order Coefficient. In future studies, this can be managed by randomizing the order of the boxes, or we could simply analyze the content of the Cards.

### B. *Threats to generalization*

To be able to generalize the findings, a sample should be drawn statistically from an intended population. With the 701 participants we do not claim that our observation can be generalized to the whole entrepreneurial population. However, given certain characteristics of the sample, we can bring implication towards early-stage entrepreneurs in Europe and Africa.

## VII. CONCLUSIONS

Entrepreneurship and Startups are essential contributors to the global economy growth, innovation and society development. Research on startups must leverage disciplines from multiple areas, i.e. entrepreneurship, management, organization and engineering. We have performed an industrial survey that collects data on how Entrepreneurs start and develop their business. We have adopted a theory of Resource and Competence (Internal View) and Environment and Market (External View) to shed light on the collected data. We are reluctant to draw too clear conclusions based on the 701 cases that we have studied. But we will point toward some indications.

In our sample, consisting of a wide range of actual entrepreneurial cases from five countries, followed between November 15[th], 2017 and March 1[st], 2018, and based on a method utilizing the Order Coefficient (measuring the order the entrepreneurs enter "Distinctions", "KeyContributions" and "KeyMarkets" into their Business Mission Board), we see indications that:

1. There is a higher tendency toward an Internal (Resource and Knowledge Based) View versus an External (Environmental Models of Competitive Advantage) View.
2. This tendency is clearer in stable economies (Europe) than in emerging economies (South Africa).
3. Cases that have a higher tendency to favour an Internal View (Resource and Knowledge Based), tend to have a higher focus on their Resources than cases favourable to an External (Environmental Models of Competitive Advantage) View. In particular, with regards to core Resources like Core Values, Visions, Owner's Objectives and Products.

Finally, we can state that an Order Based Method (using the order of the entered Business Mission elements) seems to give clearer indications on this than a Number Based (using the number of the entered Business Mission elements) Method.